\def\year{2020}\relax
\documentclass[letterpaper]{article} 
\usepackage{aaai20}  
\usepackage{times}  
\usepackage{helvet} 
\usepackage{courier}  
\usepackage[hyphens]{url}  
\usepackage{graphicx} 
\usepackage{graphicx}  
\title{Eliminating self-selection: Using data science for authentic undergraduate research in a first-year introductory course}

\author{Lior Shamir\textsuperscript{\rm 1} \\ 
\textsuperscript{\rm 1} Department of Computer Science, Kansas State University \\ 
1701D Platt St. \\
Manhattan, KS 66506 \\
lshamir@mtu.edu 
}

\begin{document}


\maketitle

\begin{abstract}
Research experience and mentoring has been identified as an effective intervention for increasing student engagement and retention in the STEM fields, with high impact on students from undeserved populations. However, one-on-one mentoring is limited by the number of available faculty, and in certain cases also by the availability of funding for stipend. One-on-one mentoring is further limited by the selection and self-selection of students. Since research positions are often competitive, they are often taken by the best-performing students. More importantly, many students who do not see themselves as the top students of their class, or do not identify themselves as researchers might not apply, and that self selection can have the highest impact on non-traditional students. To address the obstacles of scalability, selection, and self-selection, we designed a data science research experience for undergraduates as part of an introductory computer science course. Through the intervention, the students are exposed to authentic research as early as their first semester. The intervention is inclusive in the sense that all students registered to the course participate in the research, with no process of selection or self-selection. The research is focused on analytics of large text databases. Using discovery-enabling software tools, the students analyze a corpus of congressional speeches, and identify patterns of differences between democratic speeches and republican speeches, differences between speeches for and against certain bills, and differences between speeches about bills that passed and bills that did not pass. In the beginning of the research experience all student follow the same protocol and use the same data, and then each group of students work on their own research project as part of their final project of the course. Several students continued to work on the research after the semester ended, and two teams also submitted scientific papers describing their findings.
\end{abstract}



\section{Introduction}
\label{introduction}

In the recent years, undergraduate research experience has been becoming increasingly more prevalent, and different models of undergraduate research experience have been proposed and implemented \cite{russell2007benefits,PCAST12,linn2015undergraduate}. Learning through research exposes undergraduate students to educational aspects and hands-on experiences they cannot earn effectively through traditional lecture-based education \cite{hunter2007becoming}. That includes skills such as making connections among seemingly disparate pieces of information, evaluation of evidence, and bringing the requisite expertise to address complex issues \cite{AAAS09,brownell2015high}. 

In addition to its academic advantages, undergraduate research experience is an effective tool for student engagement \cite{seymour2004establishing} and consequently student retention \cite{hippel1998undergraduate,kinkel2006impact,lopatto2007undergraduate,braxton2011understanding,PCAST12}. Research experience also leads to higher grades \cite{kinkel2006impact,barlow2004making}. Undergraduate research experience was found highly effective for attracting and retaining underrepresented minority students in STEM \cite{barlow2004making,Tsu07,johnson2016recruit,collins2017undergraduate}, and the impact of undergraduate research experience on retention and graduation of underrepresented minorities is higher than the average impact of research experience on the general student population \cite{russell2007benefits,villarejo2008encouraging,jones2010importance,chemers2011role}.

While undergraduate research experience is a proven effective intervention, exposing all students to research introduces several obstacles. One-on-one mentoring is often limited by the availability of faculty who can mentor undergraduate students. In institutions that focus on undergraduate education, the number of research labs is limited, as well as the number of faculty with active research programs who can mentor undergraduate students and lead them to authentic research. That situation can be solved partially by models such as the NSF's Research Experience for Undergraduates (REU), according which students can spend a summer at a research institution. However, that model depends on the availability of funding, and therefore just a few of the students can benefit from it. More importantly, joining a research lab requires a student to actively apply and sometimes compete for the research position \cite{bangera2014course}. That process might leave many students who do not see themselves competitive or do not see themselves as researchers without access to the intervention, while sometimes rewarding the more privileged students who use these research experience opportunities to further enhance their skills \cite{tootle2019mini}. Therefore, the students who are stronger academically, have higher GPA, and see themselves as researchers are much more likely to be exposed to research experience \cite{sell2018impact,cooper2019factors}, while students who are less confident and might benefit from the intervention the most practically do not have access to the intervention \cite{salgueira2012individual}. Another obstacles is the time commitment for extracurricular activities, that might make research practically inaccessible to non-traditional students such as commuting students or students who have full-time or part-time jobs. Moreover, it has been shown that a continuous intervention is required to achieve integration of underrepresented minorities in a STEM career \cite{estrada2018longitudinal,hernandez2018undergraduate,tootle2019mini}, and therefore a time-limited research experience might not achieve the full impact of undergraduate research experience.

Here I describe a model of authentic undergraduate research experience in data science that is part of a regular first-year course. The students are exposed to the research as part of the course, and therefore all students taking the course participate in the research activities. Because the research is part of the curriculum, no selection or self-selection is applied, and no extra-curricular time commitment is needed from the students. As a first-year/first-semester introductory course, it exposes the students to research as early as possible, rather than in the junior or senior years when retention becomes less critical.

\subsection{Institutional context}
\label{institution}

Kansas State University is a public research-oriented (Carnegie classification R1) land grant university. Its undergraduate computer science program enrolls $\sim$550 students. While undergraduate research opportunities do exist, most of the research is performed by graduate students. Class size of required courses range typically from 40 students to $\sim$100 students.

The course in which the research experience is implemented is ``Introduction to Computing'', which is the most basic computer science course in the computer science degree program. It is a three credit-hour course, and the class meets twice a week. In the Fall semester of 2019, the section had 21 students enrolled, and two undergraduate teaching assistants. There is no textbook in the course, and reading assignments include papers such as \cite{leveson1993investigation} and sources that are available on-line. The course is taken normally in the first semester, with the purpose of introducing the students to a broad range of topics in computer science, and basic programming in Python. The course does not have a final exam, but students are required to submit a final paper about a topic of their choice.

\section{Educational goals}
\label{goals}

As described above, the purpose of the course is to introduce students to basic terms and topics in computer science, the history of computing, and also basic programming in Python. It briefly covers topics such as file systems, databases, boolean algebra, computer organization, programming languages, theory of computation, algorithms, information theory, and operating systems.

The purpose of the research experience is to introduce students to research methodology, which is another tool they can earn in addition to the technical and programming skills covered in the course. In the end of the course, the students are expected to be able to define a research problem, design and test solutions, perform basic statistical inference, criticize their results, and describe their work in a research paper.

As mentioned above, research experience has substantial impact on the engagement and retention of students, especially students from underrepresented minorities \cite{barlow2004making,Tsu07,collins2017undergraduate}. Therefore, exposing students to research as early as their first semester is expected to have the highest impact on retention, as retention is critical in the first year. Exposing students to research in their first semester will also provide them with the ability to join a research lab or pursue other research opportunities as early as possible, and maximize their exposure to research experience by joining a research lab or pursuing other research opportunities.

However, research lab positions are often selective, and faculty mentors often prefer to recruit students who are better prepared. Additionally, many first-year students might not see themselves competitive for earning a research position. The extra-curricular time commitment might also intimidate some of the students. Therefore, applying for a research position might not seem a high priority option for first-semester students. Also, by joining a research lab students normally work with a mentor on the mentor's research program rather than developing their own research. In that case, the students are not able to fully express their interests and identity through their research.

\section{Research experience design}
\label{design}

The research experience is done entirely as part of the course, and is designed in two phases: In the first phase all students in the course work together on the same research, and follow the same research protocol with the same data. While each group of students works independently, applying the same protocols to the same data naturally leads to the same results for all students. Although all students make the same discoveries, these discoveries are new and relevant knowledge, and are not known neither to the instructor nor to the students, and do not appear in any textbook or other existing literature. During that phase the students are introduced to the data, basic research practices, the research tools, and methods of statistical inference. That phase prepares the students to the second phase of the research experience, in which the students choose their own research problem and use the same data and same tools to make discoveries.

As first-year students, no assumptions can be made about their level of preparation, which introduces a challenge to performing authentic research activities leading to meaningful discoveries. Another challenge is that all activities need to take place as part of a course, and need to scale to a large number of students. To address these challenges, data science is used to turn existing databases into knew knowledge by using existing computational tools. That can be done without strong programming skills or other previous knowledge in computer science, and therefore suitable for freshmen level research.

The research project is 45\% of the grade. The assignments of Phase 1 are 30\% of the grade, and the paper and final presentation make another 15\% of the grade. The research activities consume a total of six meetings during the semester.

\subsection{Data}
\label{data}

While any type of data can be used, text data is selected for the research experience. Text data are normally small compared to other types of data such as image or audio. It is also often easier to use, as no strong computing facilities are typically needed to process text data, as opposed to image data where GPUs or strong processors are required to analyze images. Text can be pre-processed when needed by simple string analysis, while image or audio files are more difficult to open and read, and are therefore less suitable for research performed by first-semester students.

Many publicly accessible text databases can be used for the research experience. To further engage the students in the research, data that the students are familiar with from their personal lives should be preferred. Examples can be popular music lyrics \cite{napier2018quantitative}, and different data can be used in each semester to make different discoveries rather than repeat the exact results of previous semesters. In this course, the dataset that was selected was a corpus of several thousands congressional speeches \cite{Thomas+Pang+Lee:06a}. Each speech is labeled with the party of the speaker and their vote (for or against the discussed bill). 

The annotation of the data to democratic and republican speeches allows the identification of possible differences between republicans and democrats using data-driven discovery tools. That was done through the five steps that will be described in in the next section. Before the students start their research, a short discussion about the type of research takes place in class. The discussion provides a summary of the tools that the students will earn from the research experience such as analytical thinking, critical thinking, and the ability to make connections between different pieces of information. The purpose of the discussion is to justify the time the students spend on research, and to explain the motivation for research experience at an early stage. Due to the specific topic of the project, the students are also asked not to confuse the research findings with political views or statements that can lead to conflict or division in the classroom. The request was granted by the students, and no inappropriate political or divisive comments were made during the research experience by any of the students.

While a certain corpus was used in the semester, many other text datasets available on-line (e.g., Project Gutenberg) could have been used. As will be described in the next section, the text analysis tool is comprehensive, and the protocol can be used to analyze other datasets of text. Also, the congressional speeches used in the study were made in around 2005, and therefore the exact same protocol can be used to analyze speeches from different years.

\subsection{Phase 1: Collaborative research}
\label{phase1}

As described above, in the first phase of the research experience all students work on the same research project using the same protocol and same data, and therefore also get the same results. Because at first semester students are not experienced in research, the practice of all students doing the same research scales to a large number of students, and does not require the one-on-one attention that undergraduate research experience often requires. Using the same protocol and getting the same results also makes it much easier for the teaching assistants to grade the assignments. The research assistants just need to repeat the same experiments with the class, and compare the results of the students to their results. The research has four steps, each is summarized in a protocol and an assignment that the students need to submit. 

\begin{figure}[h]
\centering
\includegraphics[width=0.8\columnwidth]{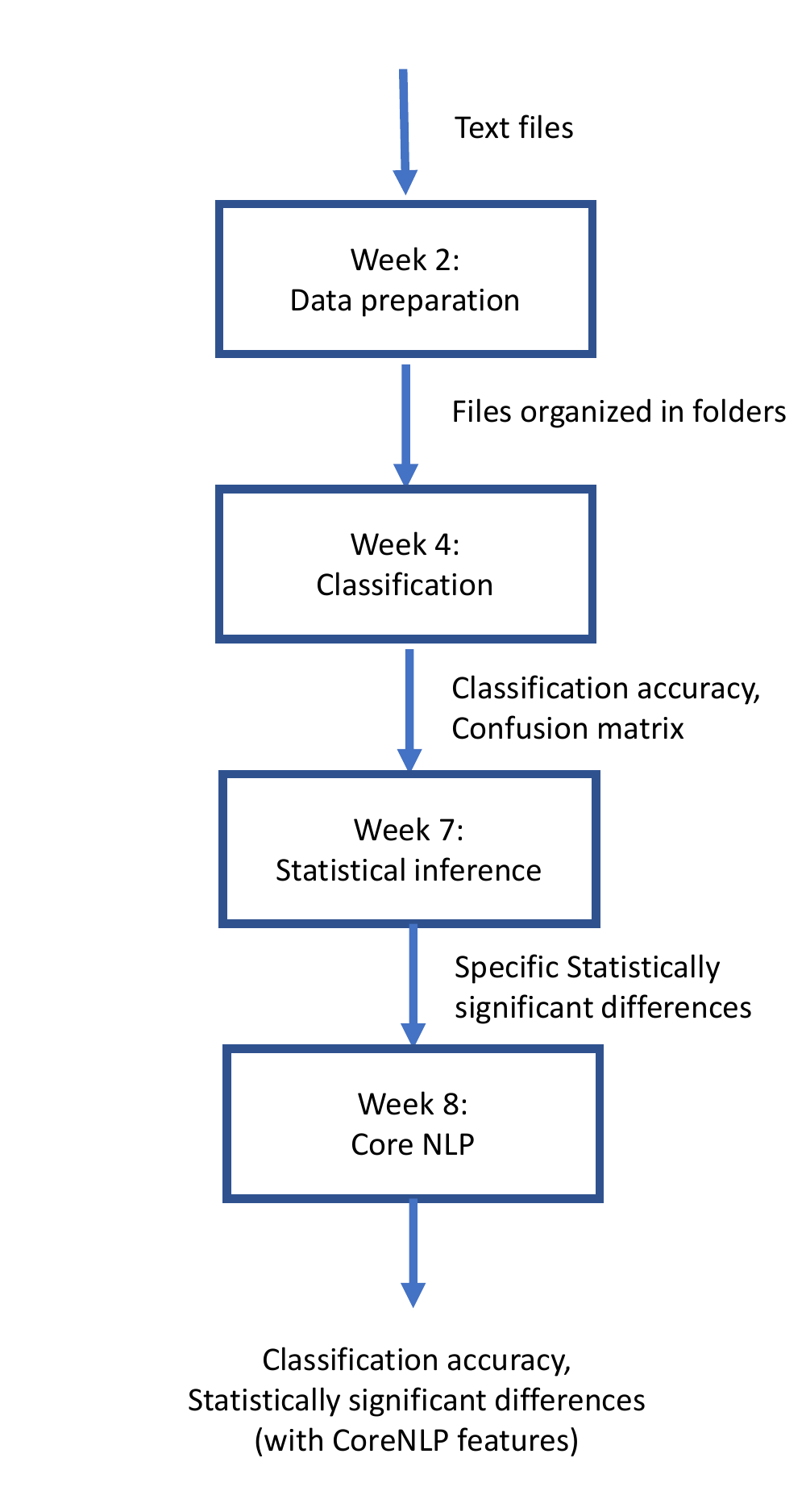} 
\caption{The four steps of the first phase of the research process. The phase starts in the second week of the semester and ends after the eighth week. During that part of the semester, all students use the same protocol, same tools, and use the same data, leading to the same results for all students.}
\label{phase1_fig}
\end{figure}

Four meetings are dedicated to the first phase of the research during the semester. The first meeting takes place in the second week of the semester, in which the research goals are described, as well as the research data that will be used in the semester. During the semester, the students work in teams of two to three students. The research process of the first phase of the undergraduate research experience is summarized by Figure~\ref{phase1_fig}, showing the different steps and the delivery of each step.

\subsubsection{Step 1: Organizing the data}
\label{step1}

The first step takes place in the second week of the semester. The students are requested to download and extract the dataset, and sort the dataset into two folders, such that one folder contains speeches made by democrat legislators and the other contains speeches made by republican legislators. The students also need to remove files that are less than 700 characters long, in order to avoid speeches that are just short comments or welcome messages, and are too short and not informative for automatic analysis. The students need to submit the number of republican and democrat speeches they have left, and their submission is graded. Through this step, the students learn about the dataset while using some basic file systems tools. 

Because the data is relatively clean, this step is not as comprehensive as typical data wrangling steps in other cases of data-driven research. On the other hand, the students do not need to spend substantial amount of time to collect and organize data, and therefore can use more of their time on the analytics part of the research.

The step consumes one meeting, in which the students are introduced to the research topic and dataset, and start working on their research in class. The students follow the instructions they are provided through an MS-Word file, which they also submit after they fill in the number of speeches that they have for each political party.

\subsubsection{Step 2: Classification and features}
\label{step2}

In the second step of the research, the students start to analyze the data. For that purpose, they use UDAT \cite{shamir2017udat}, which is an open source command-line tool designed to make discoveries in data. UDAT can be used without strong programming skills, and is freely available online also in the form of Windows binaries\footnote{http://people.cs.ksu.edu/~lshamir/downloads/udat/}. Unlike document classifiers that can just assign documents to classes, UDAT also has explainable AI aspects that provide information about the relationship between the different classes, and identify certain features that are similar or different between classes. UDAT measures the text descriptors such as readability indices, sounds of words, use of punctuation characters, use of different parts of speech, re-use of words, sentiments, and more. More information about UDAT text analysis can be found in \cite{shamir2015leveraging,alluqmani2018writing}.

By following a detailed protocol, the students use supervised machine learning to classify between democrat and republic speeches, create the confusion matrix of the classification, create the similarity matrix, use bootstrapping, and change the number of training and test samples to learn how the classification accuracy changes with different sizes of data. Additionally, the students use feature selection to identify individual text measurements that provide discriminating information between democrat and republican speeches. UDAT can provide a list of the features that have the highest discriminating power according to their LDA scores, as well as the means of the features in each class. 

The students need to follow the detailed protocol, and then submit the classification accuracy under different sizes of the training set. The students also need to provide the confusion matrix and similarity matrix that they generate using UDAT. Finally, the students need to identify the seven text measurements that have the highest discriminating power between democrat and republican speeches.

\subsubsection{Step 3: Statistical significance}
\label{step3}

In Step 2, the experiment showed that UDAT was able to identify the party of the speaker in 61\% of the cases. However, the ability to classify between republican and democrat speeches merely shows that there are differences between democrat and republican speeches, but does not identify what these differences are. The identification of discriminating features is therefore an important part in the discovery aspect of the research. 

In the third step of the process the students need to perform a basic statistical inference to determine whether the text features that were identified in Step 2 show statistically significant difference between democrats and republicans. UDAT shows the means and standard error of the means of text features measured for each class. Using that information, the students can use a statistical calculator\footnote{https://www.graphpad.com/quickcalcs/ttest1/} to compute the t-test of the difference between the mean of the feature values of the democrat speeches and the means measured for the same features in the republican speeches. Another topic is correcting the P value to multiple tests.

The mathematics of the Student t-test is not being covered in class, as the students have not yet taken calculus and are not prepared to understand the statistics, but the concept of P values is being discussed in the context of a discovery. The activity starts by describing its goals and discussing P values, and then the groups of students work in the class to determine the P values of the different features. The outcomes of the assignment that the students submit is 10 text features that have statistically significant means between democrat and republican speeches, and five text features that are not statistically significant. If no 10 text features are found to be statistically significant, the students need to mentioned that in the assignment their submit.

Through that part of the research the students discovered that democrat legislators tend to use longer words in their speeches compared to republican legislators, and the difference is statistically significant. The mean length of a word in a democrat speech was 4.72$\pm$0.006, while the mean length of a word in a republican speech was 4.64$\pm$0.007. It also showed that democrat legislators use more quotations, and use more homogeneous sounds in their selection of words, as determined by using the Soundex algorithm.

\subsubsection{Step 4: Adding CoreNLP}
\label{step4}

In the final step of the first phase the students repeat Steps 2 and Step3, but with using CoreNLP \cite{manning2014stanford}. UDAT can work with CoreNLP to identify parts of speech, as well as sentiments expressed in the speeches. After Step 2 and Step 3, the students are more experienced, and can perform that part of the research independently.

Through that step the class discovered that democrats use more nouns in their speeches, and that republican speeches express more positive sentiments than democratic speeches.

\subsection{Phase 2: Individual research}
\label{phase2}

In the second phase of the research, each team of students needs to define its own research project. That is done in the last four weeks of the semester, and provides students with the opportunity for ownership of the research, which is an important elements of undergraduate research \cite{lopatto2003essential}. That part of research is performed by the students through discussions in the classroom, and each team presents their ideas (each presentation is about 5-7 minutes) to the classroom, followed by a short discussion and comments from the instructor and the other students. In each week, the beginning of one meeting is dedicated to brief update reports from each team. The students were encouraged to use the same data, as well as the same data analysis tools, but were also given the option to use other data that is relevant to the project. 

In Phase 2, the students have sufficient knowledge about the analysis tools, and could perform simple research tasks leading to basic discoveries from data. The primary outcome of this phase is a research paper of 2000-5000 words. The students also make a short presentation about their research. The requirement to submit an original research paper replaces the previous final paper requirement of the course. That is, instead of submitting a paper that summarizes a topic of their choice, the students submit a paper about their research. Unlike the assignments submitted by the students in Phase 1, the final research papers are graded by the instructor and not by the teaching assistants.

Several students chose to use public sources of congressional speeches and obtain much larger datasets, a task that involves substantial labor not required by the course. Namely, analyzing data over a very long period of time of over 100 years of congressional speeches led to interesting insights about how legislators express themselves through speeches, and a student-authored paper written based on this study was submitted for publication in a peer-reviewed journal. Other students associated each speech file with the legislator, and identified differences between different legislators reflected by their speeches. Another example is a study by another team of students identifying differences between speeches of legislators who voted for the bill and legislators who voted against it.

\section{Results}
\label{results}

In the end of the course, all students submitted their final papers, and completed the course successfully. No student dropped the course, failed it, or avoided submitting the research outcomes. Student evaluation for the question ``Increased desire to learn about the subject'' was 4.9 (out of 5). As a first-semester course, the objective of the course is to increase student interest in computer science to engage and retain the students in the field, and therefore the student interest in the field is critical to the outcomes of the course. Fairness of grading was rated at 4.3, showing that no major concern was expressed about grading, despite the fact that 45\% of the grade was the research project. Two teams of students continued to work during winter break of 2019-2020, which led to a completion of a research paper that was submitted to a peer-reviewed journal, and another paper is in preparation.

The impact of the intervention was also tested by using a pre- and post student surveys. The survey include 15 questions adjusted from the Lopatto CURE survey \cite{denofrio2007linking}, focusing on experience and self efficacy, and measured by a forced-choice 1-5 Likert scale. The questions are the following: \newline \newline
 1. ``Even if I forget the facts, I'll still be able to use the thinking skills I learn in science"	\newline
 2. ``You can rely on scientific results to be true and correct"	\newline
 3. ``The process of writing in science is helpful for understanding scientific ideas"	\newline
 4. ``Students who do not major/concentrate in science should not have to take science courses." \newline	
 5. ``I wish science instructors would just tell us what we need to know so we can learn it."	\newline
 6. ``Creativity does not play a role in science."	 \newline
 7. ``Science is not connected to non-science fields such as history, literature, economics, or art."	 \newline
 8. ``I get personal satisfaction when I solve a scientific problem by figuring it out myself."	\newline
 9. ``I can do well in science courses.'' \newline
 10. ``Scientists know what the results of their experiments will be before they start."	 \newline
 11. ``If an experiment shows that something doesn't work, the experiment was a failure."	 \newline
 12. ``I prefer hands-on activities in the course.''	 \newline
 13. ``Sometimes in my classes I noticed unfair treatment related to race/ethnicity."	 \newline
 14. ``I prefer open-ended projects over textbook assignments."	 \newline
 15. ``The textbook is an important part of the course."	 \newline




\begin{figure}[h]
\centering
\includegraphics[width=1.1\columnwidth]{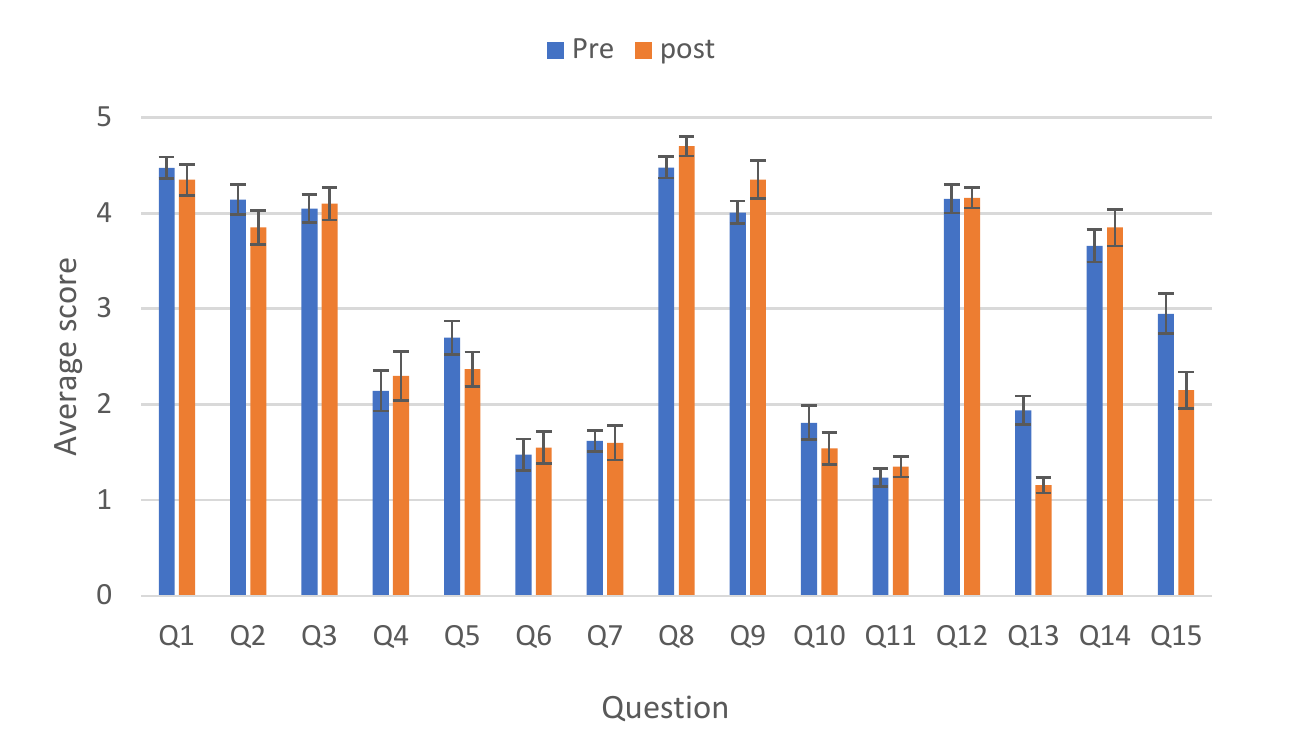} 
\caption{Average score for each of the 15 questions.}
\label{results_figure}
\end{figure}

Figure~\ref{results_figure} shows the average score of the answers of the students to each of the 15 questions. Interestingly, the question that showed the highest change between the pre and post surveys was ``Sometimes in my classes I noticed unfair treatment related to race/ethnicity''. The average answer to that question dropped from 1.94$\pm0.15$ in the pre survey to 1.15$\pm0.06$ (P$<$0.0002). Another question that showed substantial change was ``The textbook is an important part of the course". The mean of the answers to that question dropped from 2.95$\pm$0.21 in the pre survey to 2.15$\pm$0.19 in the post survey (P$<$0.0009). The textbook was not overly popular among the students in the pre survey, but its popularity decrease even further in the end of the course. The question ``I can do well in science courses'' also showed an increase from 4.01$\pm$0.11 in the pre survey to 4.35$\pm$0.12 (P$\simeq$0.04). The other questions did not show a significant difference between the pre and post surveys.


\section{Conclusion}
\label{conclusions}

Mentoring undergraduate students for research is a proven high-impact practice. However, it is difficult to scale one-on-one mentoring to a large number of undergraduate students given the typical student-faculty ratio and availability of funding. It is further limited by self-selection, as many non-traditional students do not see themselves as researchers, and might therefore not apply for research opportunities that involve mentoring. 

By using course-based undergraduate research experience all students are exposed to research, and perform research as part of the regular course that they take. That kind of research experience does not involve extra-curricular time commitment, and all students are exposed to it without a process of application or selection. It can also scale to a much larger number of students compared to the number of students that can be mentored by a single faculty using the traditional one-on-one mentoring model. 

The model proposed here uses data science foundations to make discoveries in data, and it is implemented as part of a first semester computer science course. The intervention is divided into two phases, such as first the entire class does the same research project, and then students can work in teams on their own research ideas as part of the course. 

Although the research is performed with first semester students, it leads to authentic discoveries that were not known before. Choosing a topic that the students understand such as political speeches allows the students better understand the research goals and discoveries, but also helps the students to express their own interests and identity as they work on their individual research. Other topics that can connect between data science and culture are sports, music, and art. These topics can also be linked to data science \cite{Str14,yaldo2017computational,soares2016quantitative,Geo14,george2015unsupervised,Sha12b,burcoff2017computer}, allowing students to experience data science research while expressing their culture and identity through the research. Future work will aim at expanding the research topics outside the scope of data science or artificial intelligence, as well as increasing the class size. The nature of the research activities allows scaling the research to introductory courses with larger enrollment, and therefore potentially provides a solution to the inclusion of students in undergraduate research.

\section{Acknowledgments}
This work is supported in part by NSF grant AST-1903823.

\bibliographystyle{aaai}

\bibliography{first_year_cs_cre}


\end{document}